\documentstyle[aps,prl]{revtex}
\draft
\title{Partial Densities of States, Scattering Matrices,
and Green's Functions}
\author{V. Gasparian\footnote{Permanent address:
Department of Physics, Yerevan State University, 
375049 Yerevan, Armenia}, T. Christen, and M. B\"uttiker }
\address{D\'epartement de physique th\'eorique,
Universit\'e de Gen\`eve, 24 Quai Ernest-Ansermet \\ 
CH-1211 Gen\`eve, Switzerland}
\begin{document}
\maketitle
\begin{abstract}
The response of an arbitrary scattering problem to quasi-static 
perturbations in the scattering potential is naturally expressed 
in terms of a set of local
partial densities of states and a set of sensitivities each
associated with one element of the scattering matrix.
We define the local partial densities of states and the
sensitivities in terms of functional derivatives of the 
scattering matrix and discuss their relation to the Green's function.
Certain combinations of the local partial densities of states represent
the injectivity of a scattering channel into the system and the 
emissivity into a scattering channel. It is shown that 
the injectivities and emissivities are simply related to the
absolute square of the scattering wave-function. We discuss also
the connection of the partial densities of states and the
sensitivities to characteristic times. We apply these concepts
to a $\delta $-barrier and to the local Larmor clock.   
\end{abstract}
\pacs{PACS numbers:  72.10.Bg, 03.80.+r}
%
%
%
\section{Introduction}
\label{section1}
Densities of states (DOS) play an important role in a number of different
physical contexts. For example, thermodynamic properties, tunneling
spectroscopy, electrical conduction phenomena, and charging effects depend
strongly on the DOS of the respective system under consideration. 
In recent works on ac transport in mesoscopic conductors
\cite{BTP,BU1} it was found that many results can be expressed in a
very transparent way if the concept of the DOS is generalized.
In particular, it was shown that it is not only 
the total DOS but also parts of it which have 
physical significance. In this work we point to the generality of
this decomposition of the total DOS and present expressions for
{\em partial densities of states} (PDOS) in terms of the scattering
matrix, the Green's function, and the absolute square of scattering
wave-functions. Decompositions of the total DOS into partial DOS
appear naturally in scattering problems in which one is concerned 
with the response of the system to a small perturbation
$\delta U(x)$ of the potential $U(x)$.
An example is the mentioned self-consistent treatment of electrical
ac-transport in mesoscopic conductors. Other examples are the Larmor
clock \cite{Baz,Ryb,BU2,BU3,LA,ST} and the optical clock \cite{GORC}, where
the tunneling of electrons and photons, respectively, through a
barrier containing a magnetic field is investigated. It turns out
that the PDOS determine the rotation of the spin polarization and the
Faraday rotation, respectively.\\ \indent
A one-dimensional scattering problem is characterized by a scattering
matrix with elements $s_{\alpha\beta}$. The indices $\alpha$ and
$\beta$ label out-going and incoming scattering channels,
respectively, of the system under consideration. For the two-channel
case as discussed below, these indices take
the values $1$ and $2$ to designate reference points
$x_{1}$ and $x_{2}$ at the left and the right side of the scattering
region, respectively.
The absolute squares of the scattering matrix elements determine
the transmission probability $T = |s_{21}|^{2} =|s_{12}|^{2}$
and the reflection probability $R =1-T =|s_{11}|^{2} =|s_{22}|^{2}$.
The response of the system can be characterized by a
set of {\em local} PDOS, $dn_{\alpha \beta}(x)/dE$, and a set of
{\em sensitivities}, $\eta_{\alpha \beta}(x)$, which are directly
connected to the scattering-matrix element $s_{\alpha\beta}.$
In general, the scattering matrix  
$s_{\alpha\beta}(E,U(x))$ is a function of the incident energy of 
the carriers and is a functional of the potential $U(x)$.
To linear order in a perturbation, $\delta U(x)$, 
the density response and the current response of the scattering
problem can be expressed with the help of the local PDOS
\cite{BU1}
\begin{equation}
\frac{dn_{\alpha \beta}}{dE}(x)\equiv -\frac{1}{4\pi i}
\left( s_{\alpha \beta}^{\dagger}
\frac{\delta s_{\alpha \beta}}{\delta U(x)} - 
\frac{\delta s_{\alpha \beta}^{\dagger}}{\delta U(x)}
s_{\alpha \beta}\right)
\label{lpdoss}
\end{equation}
and with the help of the sensitivities 
\begin{equation}
\eta_{\alpha \beta}(x)\equiv - \frac{1}{4\pi}
\left( s_{\alpha \beta}^{\dagger}
\frac{\delta s_{\alpha \beta}}{\delta U(x)} + 
\frac{\delta s_{\alpha \beta}^{\dagger}}{\delta U(x)}
s_{\alpha \beta}\right)\;\;,
\label{lsen}
\end{equation}
where $\delta /\delta U(x)$ denotes a functional derivative.
The local PDOS represent a decomposition of the total local
DOS \cite{BU1}
\begin{equation}
\frac{dn}{dE}(x) = \sum_{\alpha \beta}\frac{dn_{\alpha
\beta}}{dE}(x) \;\; . 
\label{ldos}
\end{equation}
They are based on both a pre-selection and
post-selection of carriers, i.e. they group carriers according to
the asymptotic region from which they arrive ($\beta$) {\em and} according
to the asymptotic region into which they are scattered ($\alpha$).
We emphasize that the PDOS are mathematical constructions. Whether
these quantities are by themselves of physical relevance might well
depend on the problem under investigation. While we find that the
off-diagonal PDOS are positive this is not always the case for the
diagonal elements. All local DOS can be obtained by summation of the
local PDOS, and the global quantities associated with an entire
segment or volume of the system are obtained by spatial
integration.\\ \indent 
It is the purpose of this work to present a discussion of the
PDOS and to relate them to the Green's
function, to dwell times, and to scattering wave-functions.
The paper is organized as follows. The scattering problem to be
considered is introduced in Sect. \ref{section2}. Some well-known
results concerning the local DOS in terms of the Green's function and
in terms of the scattering matrix are recalled in Sects. \ref{section3}
and \ref{section4}, respectively. Section \ref{section5} provides
the relation between the basic PDOS (\ref{lpdoss}) and the Green's function.
In Sect. \ref{section6}, we encounter decompositions of the local DOS
on a next higher level, based only on a pre-selection {\em} or based only
on a post-selection. We call the local PDOS which is generated by carriers
incident from the asymptotic region $\alpha$ regardless into which 
region the carriers are finally scattered the {\em injectivity}
\cite{nome} of channel $\alpha$:
\begin{equation}
\frac{d\overline{n}_{\alpha}}{dE}(x)\equiv  \sum_{\beta}\frac{
dn_{\beta\alpha}}{dE}(x)\;\;.
\label{injec}
\end{equation}
The decomposition of the local DOS into injectivities 
represents thus a pre-selection.
Similarly, we can ask about a decomposition of the local DOS 
into {\em emissivities}
according to the asymptotic region into which carriers are scattered
regardless of the channel through which the carriers 
entered the scattering region. The emissivity into channel 
$\alpha$ is given by  
\begin{equation}
\frac{d\underline{n}_{\alpha}}{dE}(x)
= \sum_{\beta}\frac{dn_{\alpha \beta}}{dE}(x)\;\;.
\label{emiss}
\end{equation}
The decomposition of the local DOS into emissivities 
represents thus a post-selection.\\ \indent
In Sect. \ref{section7} we relate the injectivity (and the
emissivity) to the time a particle dwells in a narrow region.
The dwell time \cite{BU2,BU3,LA} is in turn connected to the
absolute square of the scattering wave-function.
For instance, Eq. (\ref{injec}) has a simple interpretation in terms
of the time a carrier is {\it dwelling} in an interval $dx$ regardless
of where it is finally scattered. In terms of the scattering
wave-function $\Psi_{\alpha,B}(x)$ which has unit incident amplitude in the
region $\alpha$ in the presence of a uniform magnetic field $B$,
the dwell time of a particle in an interval $dx$ at the point
$x$ is $d\tau_{\alpha,B} = dx |\Psi_{\alpha,B}|^{2}/J$ where $J$ is the 
incoming current carried by the state $\Psi_{\alpha,B}$.
We show that the injectivity is directly related to the dwell time
and, therefore, to the wave function according to the expression
\begin{equation}
d\tau_{\alpha,B} (x) =
h \frac{d\overline{n}_{\alpha,B}}{dE}(x) \; dx =
 \frac{|\Psi_{\alpha,B}(x)|^{2}}{J} \; dx \;\;.
\label{injedw}
\end{equation}
Similarly, the emissivity is related to the square of the amplitude of the 
wave function $\Psi_{\alpha,-B}(x)$ calculated for a reversed magnetic
field: 
\begin{equation}
d \tau_{\alpha,-B}(x) = 
h \frac{d\underline{n}_{\alpha,B}}{dE}(x) \; dx =
\frac{|\Psi_{\alpha,-B}(x)|^{2}}{J} \; dx \;\;.
\label{emisdw}
\end{equation}
The injectivity and the emissivity are identical
in the absence of a magnetic field. A connection between the functional
derivatives of the scattering matrix and the local absolute squares
of the wave functions is obtained from a combination of Eqs. (\ref{lpdoss})
and (\ref{injedw}).\\ \indent
Section \ref{section8} is devoted to the sensitivities (\ref{lsen}),
which can be understood as local response of the transmission probabilities
to a potential change. They are of great importance in a self-consistent
theory of nonlinear conduction \cite{BC1}. Finally, we discuss in
Sect. \ref{section9} two examples, namely a localized impurity in a
one-dimensional conduction channel and the local Larmor clock.\\ \indent
Before proceeding, we mention that 
there are some recent experimental indications for the relevance
of the PDOS. In an impressive experiment with a quantum Hall system,
Chen et al. \cite{CSBS} measured capacitance coefficients in a 
three-terminal geometry and presented an interpretation in terms of 
PDOS. Christen and B\"uttiker \cite{CB1} discussed the low-frequency 
admittance of quantized Hall conductors with arbitrary
topologies. The same authors found that the dependence of the
properties of a quantum point contact on the PDOS results in steps of the
capacitance and of the low-frequency admittance in synchronism
with the conductance steps \cite{CB2}. Leadbeater and Lambert
\cite{LL} explained that an experimentally found \cite{MREWF}
asymmetry in the STM tunneling conductance into vortices in
a superconductor is due to an asymmetry in the injectivity associated
with particle/hole channels.
\section{The scattering problem}
\label{section2}
Consider the one-dimensional scattering problem sketched in
Fig. \ref{fig1}. In a region $x_{1}<x<x_{2}$
scattering is assumed to be purely elastic.
Of interest are the PDOS and the sensitivities in this region.
The global scattering properties are described
by the scattering-matrix elements $s_{\alpha \beta}$ which are
the ratio of the current amplitudes of the out-going waves at
$x_{\alpha}$ and of the incoming waves at $x_{\beta }$.
Note that the scattering matrix is here defined with respect to current
amplitudes at {\em finite} $x_{\alpha}$ ($\alpha = 1,2$), and not
with respect $x\to \pm \infty $. In the sequel, all quantities are
absolute quantities rather than defined relatively to a free
particle. For example, we deal with scattering phases rather than phase
shifts and with total times rather than delay times (i.e. time
differences relative to the free particle).\\ \indent
Scattering is due to a
stationary potential $U(x)$ localized in the region $[a,b]$. To be
definite, we assume $x_{1}<a<0<b<x_{2}$ and that an absolute maximum
of the potential, if any, is located at $x=0$.
In the regions $\Omega_{1} = [x_{1},a] $ and
$\Omega _{2}=  [b,x_{2}]$ the potentials are
constant and take the values $U_{1}$ and $U_{2}$, respectively.
Quantities associated with these regions are labeled by roman
indices in contrast to the greek labels which designates the
boundaries ($x_{\alpha}$) of the system
(for, e.g., an electrical conductor greek and
roman labels designate contacts and regions in the conductor,
respectively). This distinction is conceptually important.
For example, in general the number of contacts differs from the number of
relevant regions \cite{CB1}.\\ \indent
In the region $\Omega _{l}$ the wavenumber $k_{l}$
of a particle is related to the energy $E$ by $E = (\hbar k_{l})^{2}/2m
+U_{l}$. This energy dispersion defines in, e.g., $\Omega _{1}$ a
scattering channel with a left
incoming branch in which particles have the positive velocity
$v_{1} = \hbar k_{1}/m$ and an out-going branch 
in which particles have the negative velocity $ -v_{1}$. An
analogous (inverse) relation holds in $\Omega _{2}$.\\ \indent
Below, the classical turning points
$x_{L}$ and $ x_{R}$ at the left and the right side of an opaque
barrier, respectively, will be important. For a transparent
scattering obstacle, where
${\rm Max}_{x}\{ U(x)\}<E$, we define $x_{L}=x_{R}=0$ at the maximum
of the barrier. We call the system `large', if
$x_{L}-x_{1} $ or $ x_{2}-x_{R}$ is much larger then the typical
wavelength $\lambda $ of the particle. The semiclassical (WKB)
regime is applicable, if the characteristic scale of the space dependence
of $U(x)$ is much larger then the typical wave length $\lambda $.
\section{Density of States and Green's function}
\label{section3}
We recall briefly some useful results concerning the retarded 
single-particle Green's function $G(x,\tilde x)$ \cite{revgreens}.
The scattering problem defined in the last section is associated with
the Hamiltonian 
\begin{equation}
H=-\frac{\hbar ^{2}}{2m}\partial _{x}^{2} + U(x)\;\;.
\label{hamilton}
\label{equation}
\end{equation}
The retarded Green's function is then defined as the regular solution of
\begin{equation}
\left( E+i\epsilon -H \right)
G(x,\tilde x) =
\delta (x-\tilde x)\;\;,
\label{greens}
\end{equation}
where one takes the limit $\epsilon \to 0^{+}$ and
where $\delta (x)$ is the Dirac delta function. This Green's function can
be interpreted as the quantum mechanical probability amplitude for
the propagation of the particle from $\tilde x $ to $x$.
Since we are interested only in scattering states we concentrate
on the continuous part of the spectrum of $H$.
Effects of the discrete part of the
spectrum (belonging to localized states) on the PDOS are
disregarded.\\ \indent 
The local DOS at an energy $E$ is given by the imaginary part of the
Green's function
\begin{equation}
\frac{dn}{dE}(x)= - \frac{1}{\pi} {\rm Im} \{ G(x,x) \} \;\;.
\label{ldosg}
\end{equation}
From the local DOS one obtains the global DOS 
\begin{equation}
\frac{dN}{dE}= \int _{x_{1}}^{x_{2}}
dx\: \frac{dn}{dE}(x) = - \int _{x_{1}}^{x_{2}}
dx\: \frac{1}{\pi} {\rm Im} \{ G(x,x) \} \;\;.
\label{tdosg}
\end{equation}
For example, the retarded Green's function of a free particle
($U(x)\equiv 0$) with wavenumber $k=\sqrt{2mE}/\hbar$ is given by
$G(x,\tilde x)= -i/(\hbar v)\exp (ik |x-\tilde x|)$
with the particle velocity $v=\hbar k/m$. Consequently,
the local DOS of a ballistic channel is $2/hv$. The
DOS in a channel of length $L$ is then $2L/hv$.
This can be seen by noticing that $kL$ gives the number $N(E)$ of
nodes (which count the states) of the wave function with energy $E$;
the derivative with respect to energy yields the DOS.
\section{Density of States and Scattering Matrix}
\label{section4}
Some useful results which relate the DOS to the scattering matrix follow
already from general expressions of the scattering matrix elements
in terms of the transmission and reflection probabilities and the scattering
phases  $\phi $ and $\phi \pm \phi _{a}.$ Here $\phi $ is the total phase
accumulated in a transmission event and $\phi \pm \phi _{a}$ are the phases
accumulated by a particle which is incident from the left or the right and 
which is reflected. The phases are measured at $x_{1}$ and $x_{2}$. 
The scattering-matrix elements can be written in the form 
\begin{equation}
{\bf s}(E) = \left( \begin{array}{cc}
r & t \\
t & r^{\prime} \\
\end{array} \right) 
= \left( \begin{array}{cc}
-i \sqrt{R}\exp (i\phi + i\phi _{a})  & \sqrt{T} \exp (i\phi)  \\
\sqrt{T} \exp (i\phi) & -i \sqrt{R}\exp (i\phi - i\phi _{a}) \\
\end{array} \right)  \;\;.
\label{scatmat}
\end{equation}
This scattering matrix is assumed to be symmetric
which holds in the absence of a magnetic field. For a spatially
symmetric barrier and for symmetrically located $x_{1}$ and
$x_{2}$ the phase asymmetry $\phi _{a}$ vanishes and one has
additionally $r=r^{\prime}$.\\ \indent
Avishai and Band \cite{AB} showed that the one-dimensional DOS of
a large system is given by the energy derivative of the
scattering phase
\begin{equation}
\frac{dN}{dE}= \frac{1}{4\pi i} \sum_{\alpha \beta}\left(s_{\alpha
\beta}^{\dagger }\frac{ds_{\alpha \beta} }{dE} -
\frac{ds_{\alpha \beta}^{\dagger } }{dE}s_{\alpha \beta}\right) =
\frac{1}{\pi} \: \frac{d\phi }{dE}\;\;.
\label{tdoss}
\end{equation} 
In the absence of scattering the phase is given by $\phi =kL$ which
again implies $dN/dE=2L/hv$. However, the DOS of
an {\em open} and {\em finite} system is not given by
Eq. (\ref{tdoss}) but must be calculated by
the spatial integration (\ref{tdosg}) of the local DOS.
Gasparian et al. \cite{GORCP} calculated the integral (\ref{tdosg})
in a finite region and expressed the final results in terms
of the scattering-matrix elements. The integrated DOS differs
from (\ref{tdoss}) by a correction
which contains the reflection amplitudes divided by the energy:
\begin{equation}
\frac{dN}{dE}= \frac{1}{\pi} \frac{d \phi }{d E}+
{\rm Im} \{ \frac {r + r^{\prime}}{4\pi E}\} =
\frac{1}{\pi}\left( \frac{d\phi}{dE}-\frac{\sqrt{R}}{4\pi E}\cos
(\phi)\cos(\phi _{a}) \right)
\;\;. 
\label{vdos}
\end{equation}
The relative difference of the results (\ref{tdoss}) and (\ref{vdos})
is of ${\cal O}(\lambda /L)$. This implies that
the correction term can be neglected for large systems,
for large energies, and in the semiclassical (WKB) case (and, of
course, if $R$ is negligible).
The local DOS can not only be obtained from Eq. (\ref{ldosg}),
but alternatively also from Eq. (\ref{ldos}), i.e. from the scattering
matrix $s_{\alpha \beta}$ and its functional derivative with respect
to the potential $U(x)$. In the next section we work out a relation
between these two approaches in view of the PDOS.
\section{Partial densities of states}
\label{section5}
The aim now is to derive simple expressions for the basic PDOS
(\ref{lpdoss}) by using the Green's function. Before doing this,
we discuss two cases which yield some vivid insight. First,
we calculate the PDOS in a large system directly from the energy 
derivative of the scattering matrix (\ref{scatmat}). Secondly, we
construct the local PDOS in the WKB approximation from
phase-space arguments.   
\subsection{Global partial densities of states in a large interval}  
\label{section51}
If we are not concerned with effects of the order of ${\cal O}(\lambda /L)$
the global PDOS
in a large interval can be found by taking energy derivatives of the 
scattering matrix
\begin{equation}
\frac{dN_{\alpha \beta}}{dE}= \frac{1}{4\pi i} \left(s_{\alpha
\beta}^{\dagger }\frac{ds_{\alpha \beta} }{dE} -
\frac{ds_{\alpha \beta}^{\dagger } }{dE}s_{\alpha \beta}\right)\;\;. 
\label{pdoss}
\end{equation} 
Using the specific from (\ref{scatmat}) of  the scattering matrix
one finds
\begin{eqnarray}
\frac{dN_{11}}{dE} & = & \frac{R}{2\pi} \frac{d(\phi +\phi_{a})}{dE} 
\label{pdoss11} \\
\frac{dN_{12}}{dE} & = & \frac{dN_{21}}{dE}=
\frac{T}{2\pi} \frac{d\:\phi}{dE}  
\label{pdoss12} \\
\frac{dN_{22}}{dE} & = & \frac{R}{2\pi} \frac{d(\phi -\phi_{a})}{dE} 
\;\;.
\label{pdoss22} 
\end{eqnarray} 
Note that the local quantities $dn_{\alpha \beta}/dE$ can be written
formally by replacing the derivatives $d/dE$ in
(\ref{pdoss11})-(\ref{pdoss22}) by the functional
derivatives $-\delta / \delta U(x)$.
Clearly, the dependence of the PDOS (\ref{pdoss11})-(\ref{pdoss22}) on
the transmission and reflection probabilities had to be expected.
The PDOS $dN_{12}/dE$ associated
with particles transmitted from the right to the left is $T/2$ times
the total DOS, while $dN_{11}/dE$ must be proportional to the
reflection probability. In the following paragraph such arguments are
used to construct the local PDOS in the semiclassical approximation.
\subsection{Semiclassical partial densities of states}
\label{section52}
In the semiclassical case, the local PDOS can be obtained with the
help of the simple phase-space arguments of Ref. \cite{CB2}.
In Fig. \ref{fig2} we sketched the classical phase space of the
scattering problem of Fig. \ref{fig1}. Consider the
trajectories at energy $E$ (thin curves in Fig. \ref{fig2}).
Since the phase-space area per state corresponds to Planck's
constant $h$, the semiclassical DOS is related to the energy
derivative of the phase-space area $ \Phi $ enclosed
by the trajectories of positive and negative momentum:
$dN^{(qc)}/dE = h^{-1} d\Phi /dE$
\cite{LaLi}. In Fig. \ref{fig2}, the phase-space region $d\Phi$
is indicated by a grey filling. From the classical equation of motion
one obtains for the local DOS $ dn^{(qc)}/dE
=(2/hv)(1-U(x)/E)^{-1/2}$ and $ dn^{(qc)}/dE =0$
for real and imaginary momenta, respectively.\\ \indent
We construct now the local PDOS from the local DOS and the
reflection and transmission probabilities.
We mention that there exists a WKB expression for $T$ in both
cases of opaque and transparent barrier \cite{MG}. We may restrict
ourselves to $x_{1}<x<x_{L}$; for $x_{R}<x<x_{2}$ the results are obtained
by appropriately interchanging the indices $1$ and $2$.
Since a relative fraction $T$ of the particles
with positive momentum are transmitted from $x_{1}$ to $x_{2}$
(and vice versa) one has
\begin{equation}
\frac{dn_{12}^{(qc)}}{dE}(x)= \frac{dn_{21}^{(qc)}}{dE}(x)=
\frac{T}{2} \frac{dn^{(qc)}}{dE}(x)  \;\;.
\label{pdoswkb12}
\end{equation}
A relative fraction $R$ of the particles with positive momentum
and a relative fraction $1-T$ of particles with negative momentum
contribute to the local PDOS of reflected particles, hence
\begin{equation}
\frac{dn_{11}^{(qc)}}{dE}(x)= R \frac{dn^{(qc)}}{dE}(x)\;\;.
\label{pdoswkb11}
\end{equation}
Because on the left of the barrier there are no classical
trajectories which both emanate at and return to $x_{2}$, one
concludes
\begin{equation}
\frac{dn_{22}^{(qc)}}{dE}(x)=0 \;\;.
\label{pdoswkb22}
\end{equation}
Next, we calculate the quantum
mechanical corrections to these expressions. Interestingly, it turns
out that for the fully quantum mechanical problem $dn_{22}/dE$
does not vanish.  
\subsection{Partial densities of states and the Green's function}
To derive exact expressions for the local PDOS, we start from the
Fisher-Lee relation \cite{FL} 
between the scattering matrix and the Green's function
\begin{equation}
s_{\alpha \beta}=
-\delta _{\alpha \beta} + i \hbar \sqrt{v_{\alpha} v_{\beta}}
\: G(x_{\alpha}, x_{\beta}) \;\;.
\label{fisherlee}
\end{equation}
Insertion of Eq. (\ref{fisherlee}) in Eq. (\ref{lpdoss}) gives
\begin{equation}
\frac{dn_{\alpha \beta}}{dE}(x) = 
- \frac{ \hbar \sqrt{v_{\alpha}v_{\beta}}}{4 \pi}
\left( s^{\ast}_{\alpha \beta}\:
\frac{\delta G (x_{\alpha},x_{ \beta})}{\delta U(x)} \; + \; 
{\rm h.c.} \right)\;\;,
\label{lpdosfl}
\end{equation} 
where the asterisk indicates complex conjugation. The functional
derivative of the Green's function $\delta G/\delta U$
is calculated by adding to the Hamiltonian (\ref{hamilton})
the local potential variation $\delta U(x)= \delta U_{0} \: 
\delta (x-x_{0}) $. One finds for the variation of the Green's function
$ \delta G(x,\tilde x)(x_{0}) = \delta U \: G(x,x_{0})G(x_{0},\tilde
x)$ which implies   
\begin{equation}
\frac{\delta G (x_{\alpha},x_{ \beta}) }{\delta U(x)}=
G (x_{\alpha},x )G (x,x_{ \beta}) \;\;. 
\label{varigreens}
\end{equation}
Equation (\ref{lpdosfl}) can thus be written as 
\begin{equation}
\frac{dn_{\alpha \beta}}{dE}(x)=
- \frac{ \hbar \sqrt{v_{\alpha}v_{\beta}}}{4 \pi}
\left( \; s^{\ast}_{\alpha \beta}\:
G (x_{\alpha},x )G (x,x_{ \beta}) \; + \; {\rm h.c.} \; \right)\;\;.
\label{lpdosv}
\end{equation}
This formula represents the central result of our work. Together with
Eq. (\ref{fisherlee}), it expresses the local PDOS fully in terms of
the Green's function. For certain cases,
Eq. (\ref{lpdosv}) can be transformed to simpler expressions containing
only $T$, $v_{\alpha}$, and $dn/dE$.
It is shown in the appendix that the PDOS of transmitted
particles is   
\begin{equation}
\frac{dn _{12}}{dE} (x)= \frac{dn_{21}}{dE}(x)= 
\frac{T}{2} \: \frac{dn}{dE}(x) 
\;\;,
\label{lpdos12}
\end{equation}
which has the same form as the the WKB result (\ref{pdoswkb12}).
Since the sum over all local PDOS equals the local DOS it holds 
\begin{equation}
\frac{dn_{11}}{dE}(x)+ \frac{dn_{22}}{dE}(x)= R\frac{dn}{dE}(x)\;\;.
\label{sumrule1}
\end{equation}
Unfortunately, it is not possible in general to find expressions
for $dn_{\alpha \alpha}/dE$ which are similarly simple as Eq.
(\ref{lpdos12}). In the flat potential regions, however, it is
possible. Consider, e.g., $x\in \Omega _{1} $. The results for $\Omega _{2}$
follow from an appropriate exchange of the indices.
We find for the local PDOS in $\Omega _{1}$ (see the appendix) 
\begin{eqnarray}
\frac{dn _{11}}{dE} (x) & = &
R\frac{dn}{dE}(x) -\frac{dn_{22}}{dE}(x)
\label{lpdos11} \\
\frac{dn_{22}}{dE} (x) & = &
\frac{T}{2}\left(\frac{2}{hv_{1}}-\frac{dn}{dE}(x)\right)
\label{lpdos22}
\end{eqnarray} 
The WKB results, Eqs. (\ref{pdoswkb12})-(\ref{pdoswkb22}),
are immediately recovered if one recalls that
$dn^{(qc)}/dx=2/(hv_{l})$ in $\Omega _{l}$.
Clearly, for vanishing transmission $T =0$ there are no states on the
left side of a barrier which are scattered from $x_{2}$ to $x_{2}$,
and (\ref{lpdos11}) and (\ref{lpdos22}) vanishes.
Below we show that for finite $T$ the local PDOS
(\ref{lpdos22}) can be negative. Thus one concludes that, in general,
the basic PDOS can not be interpreted as densities of
states in the usual sense of the word. 
\section{Injectivity and emissivity}
\label{section6}
In the introduction we mentioned that in many cases the injectivity
(\ref{injec}) and the emissivity (\ref{emiss}) are the physically
relevant PDOS. From their definition it follows that the sum
over all emissivities and the sum over all injectivities
is equal to the local DOS. In the absence of a magnetic field, as is
the case here, injectivity and emissivity are equal to each other
\cite{BU1}. In principle, they must be calculated from Eq.
(\ref{lpdosv}). Again, they simplify considerably in the regions where
the potential is uniform. From Eqs. (\ref{lpdos12})-(\ref{lpdos22}) one
obtains in, e.g., $\Omega _{1}$
\begin{eqnarray}
\frac{d\overline{n}_{1}}{dE}(x)=
\frac{d\underline{n}_{1}}{dE}(x)  & = &
\frac{dn}{dE}(x) -\frac{T}{hv_{1}}
\label{emis1} \\
\frac{d\overline{n}_{2}}{dE}(x) = \frac{d\underline{n}_{2}}{dE}(x) & = &
\frac{T}{hv_{1}}   
\label{emis2} \;\;.
\end{eqnarray}
The injectivity $d\overline{n}_{2}/dE$ is constant in $\Omega _{1}$.
This follows from the fact that the injectivity is proportional
to the absolute square of the scattering wave-function which is shown
in the next section.\\ \indent
The injectances and emittances of a large system can be expressed by
energy derivatives of the scattering-matrix elements
\begin{eqnarray}
\frac{d\overline{N}_{1}}{dE}=\frac{d\underline{N}_{1}}{dE}  & = &
\frac{1}{2\pi} (\frac{d\phi}{dE} + R \frac{d \phi_{a}}{dE})   
\label{emit1} \\
\frac{d\overline{N}_{2}}{dE}= \frac{d\underline{N}_{2}}{dE} & = &
\frac{1}{2\pi} (\frac{d\phi}{dE} - R \frac{d \phi_{a}}{dE})   
\label{emit2} \;\;.
\end{eqnarray}
The injectance (and emittance) contains the reflected part
of the DOS associated with the phase asymmetry. In the following
section we show that the injectance (and the emittance) is related to
a dwell time.     
\section{Dwell times}
\label{section7}
There exists a vast literature on characteristic times
(e.g., traversal, reflection, and dwell times) 
for the motion of a particle in the presence of a barrier potential
(see, e.g., Refs. \cite{LM,HS} and Refs. therein).
Such times are closely related to PDOS. In classical mechanics 
the time $\tau _{cl} $ needed by a particle to traverse
a piece of a trajectory 
is given by the energy derivative of the phase-space area 
enclosed by the trajectory and the space axis. If one recalls the
relation between phase-space area and DOS one finds for
the time the particle resides in a given
region and the semiclassical DOS a relation of the type
$\tau _{cl}= h \: dN/dE$.\\ \indent
The definition of characteristic times in quantum dynamics is a more
subtle undertaking. We derive now an expression for the
time $d\tau _{\alpha} $ a particle injected at $x_{\alpha} $
dwells in the region $[x,x+dx]$, and we show that it
is related to the injectivity $d\overline{n}_{\alpha}/dE$ at $x$
\cite{BU3,BC1}.
Assume that a particle current $J$ is injected at, e.g., $x_{1}$.
The dwell time in a neighbourhood of $x_{0}$ is defined as the ratio
of the particle number in the interval $[x_{0},x_{0}+dx]$ and the
incoming current:
\begin{equation}
d\tau _{1}(x_{0}) = \frac{|\Psi (x_{0})|^{2}}{J}\: dx \;\;.
\label{dwellt}
\end{equation}
Obviously, this equation describes a balance equation:
the injected current equals the decay rate of the probability in
$[x_{0},x_{0}+dx]$. To calculate the dwell time, we follow closely
Ref. \cite{BU3} and introduce an infinitesimal
particle absorption of strength
$d\Gamma $ at $x_{0}$. This absorption
is described by an imaginary perturbation
$dH = -i(\hbar/2) \delta
(x-x_{0})\: d\Gamma $ of the Hamiltonian (\ref{hamilton}).
The perturbed Hamiltonian is not hermitian which implies that
the continuity equation for the quantum mechanical
probability density obtains a sink term at $x_{0}$:
\begin{equation}
\partial _{t} |\Psi (x)|^{2} + \partial _{x}j = -d\Gamma |\Psi
(x)|^{2}\delta(x-x_{0}) \;\;.  
\label{continuity}
\end{equation} 
Here, $j$ is the usual quantum mechanical current density. 
The current of absorbed particles is $dj=-d\Gamma |\Psi (x_{0})|^{2}$.
Thus, the infinitesimal dwell time can be written as
$d\tau _{1}=-dx (dj/d\Gamma)/J$. Now we use $dj=J(dR+dT)$,
where $dR$ and $dT$ are the variations of the
reflection and the transmission probabilities,
$R=|s_{11}|^{2}$ and $T=|s_{21}|^{2}$, respectively, for 
particles coming from the left. These variations
are obtained from an expansion
of the scattering matrix elements, $s_{\alpha 1}=s_{\alpha
1}^{(0)} +(\delta s_{\alpha 1}/\delta U)\:d H $. With the help of Eq.
(\ref{lpdoss}) one can write $dR= -hd\Gamma (dn_{11}(x_{0})/dE)$ and
$dT= -hd\Gamma (dn_{21}(x_{0})/dE)$. The case where particles are
injected at $x_{2}$ is treated analogously. Dropping the index of
$x_{0}$, the dwell time in a region $[x,x+dx]$ for
particles coming from $x_{\alpha}$ can be expressed in the form
\begin{equation}
d\tau _{\alpha}(x) = h \frac{d\overline{n}_{\alpha}}{dE}(x)\: dx
\label{dwelltime}
\end{equation} 
which is proportional to the injectivity.
The dwell time $\tau _{\alpha}$ of a finite region is obtained with
a spatial integration, i.e. it is essentially the injectance
$d\overline{N}_{\alpha}/dE$ of this
region.  The specific dependence on the magnetic field stated in Eqs.
(\ref{injedw}) and (\ref{emisdw}) is a consequence of
reciprocity, i.e. $d\overline{n}_{\alpha, B}/dE =
d\underline{n}_{\alpha ,-B}/dE$  \cite{BU1}. The characteristic
times associated with finite
regions must be calculated by spatial integration of a density
(the injectivity) and are not given simply by energy
derivatives of phases. This was already clear
in the discussion of the collision times by Smith \cite{S60}
by Jauch and Marchand \cite{JM}. 
It has also been pointed out by Gasparian and Pollak \cite{GP}
when they compared Larmor-clock times with times derived from energy
derivatives. Recently, a relation between DOS and dwell times has
been investigated by Iannaccone \cite{IA}. To which extent the times
obtained from energy derivatives
provide a reasonable approximation follows from the remark at the
end of Sect. \ref{section4}. In particular, for large systems this
approximation can be accepted.
\section{Sensitivities}
\label{section8}
Let us next investigate the sensitivities (\ref{lsen}) and their
connection to the Green's function. We first mention that the sensitivities
are simply related to the functional derivatives of the 
transmission probability. We have $4\pi \eta _{12}=-
\delta T/\delta U(x)$. The unitarity of the scattering matrix implies
immediately $\eta _{12} $ $= \eta _{21} $ $= -\eta _{11} $
$= -\eta _{22}$ $\equiv \eta $. In the present case of a two-channel
scatterer the sensitivities are characterized by a single quantity
$\eta$ which describes the dependence of the transmission probability
on the local potential. Along the lines of Sect. \ref{section5}
one derives (see appendix)
\begin{equation}
4 \pi \eta = -\frac{\delta T}{\delta U(x)} = 
-2 T \: {\rm Re} \{ G(x,x)\}\;\;.
\label{sens1}
\end{equation}
This result states that the real part of the diagonal elements  
of the Green's function is essentially the sensitivity.
Together with Eq. (\ref{ldosg}), this leads to an expression
for the diagonal elements of the Green's function  
\begin{equation}
G(x,x) = -\frac{2\pi }{T}\eta(x)- i \pi \frac{dn}{dE}(x)=
\frac{\delta}{\delta U(x)}(\ln \sqrt{T}+i\pi N)  \;\;.
\label{shortgreens}
\end{equation} 
The sensitivity plays a role which is 
complementary to that of the local DOS. We mention
that from the knowledge
of the sensitivity and the DOS, one can not only derive the diagonal
elements (\ref{shortgreens}) of the Green's function, but in principle
also the non-diagonal elements. This follows immediately from
Eq. (\ref{app1}) in the appendix. 
\section{Examples}
\label{section9}
In this section we present two examples. First, we consider
a channel with a delta barrier which describes,
e.g., a one-dimensional conductor with a localized impurity.
Secondly, we discuss the local Larmor clock which turns
out to be ultimately related to the local PDOS.
\subsection{The delta-barrier}
\label{deltabarrier}
As a simple example, consider a ballistic conductor ($U_{1}=U_{2}=0$)
containing a delta-function impurity $U(x)=V\delta (x)$ with
$V\geq 0$. For
convenience, we introduce the dimensionless
quantity $w=V/(\hbar v)$, where $v$ is
the particle velocity. The local PDOS can be calculated either
directly by introducing a further $\delta $-potential of
infinitesimal strength $\delta U$,
calculating the scattering matrix from the
transfer matrix, and using the definition (\ref{lpdoss}). Or it can
be calculated with the help of the diagonal elements of the Green's
function and using the results derived above. The diagonal elements
of the retarded Green's function are given by
\begin{equation}
G(x,x)=-\frac{i}{\hbar v} \left(1-\frac
{w(i+w)}{1+ w^{2}}(\cos {2kx} + i\sin {2k|x|})\right) \;\;, 
\label{deltagreens}
\end{equation}  
and the transmission and reflection probabilities are $T=1/(1+w^{2})$
and $R=w^{2}/(1+w^{2})$, respectively. 
With the help of the function $f(x) = w\cos(2kx)- \sin(2k|x|)$
we can write the PDOS in the region $x_{1}<x<0$ in the form
\begin{eqnarray}
\frac{dn_{11}}{dE}(x) & = & \frac{R}{hv}\left(2-
\frac{1}{w}\frac{1+2w^{2}}{1+w^{2}} f(x) \right)\;\;,
\label{delta11} \\
\frac{dn_{12}}{dE}(x) & = & \frac{dn_{12}}{dE}(x) =
\frac{T}{hv}\left(1-\frac{w}{1+w^{2}}f(x) \right)\;\;,
\label{delta12} \\
\frac{dn_{22}}{dE}(x) & = & \frac{T}{hv}
\frac{w}{1+w^{2}} f(x) \;\;.
\label{delta22}
\end{eqnarray}
Note that $f(x) $ contains fast Friedel-like oscillations. In particular,
$dn_{22}/dE$ which vanishes in a semiclassical consideration
contains oscillating quantum-mechanical correction terms.
Using $w/(1+w^{2})=\sqrt{RT}$, the injectivities (and the emissivities)
can be written as
\begin{eqnarray}
\frac{d\overline{n}_{1}}{dE}(x) & = &
\frac{2-T}{hv} -\frac{2}{hv}\sqrt{RT}\:f(x) \;\;,
\label{demis1} \\
\frac{d\overline{n}_{2}}{dE}(x) & = & \frac{T}{hv}   
\label{demis2} \;\;.
\end{eqnarray}
The local DOS is
\begin{equation}
\frac{dn}{dE}(x) = \frac{2}{hv}\left(1-\sqrt{RT}f(x)\right)\;\;.
\label{deltaldos}
\end{equation}
The local PDOS $dn_{\alpha \beta}/dE$ and the local DOS
$dn/dE$ are plotted in Fig. \ref{fig3}.
Note that the local PDOS $dn_{\alpha \alpha}/dE$ associated
with reflection can be negative. The injectivity $d\overline{n}_{1}/dE$
is shown in Fig. \ref{fig4}. This injectivity is proportional to
the absolute square of the scattering wave-function and is, at the
right side of the barrier, proportional to $T$ and space
independent.\\ \indent 
For a symmetric system ($x_{2}=-x_{1}=L/2$), the global DOS is
\begin{equation}
\frac{dN}{dE} = \frac{2L}{hv} + \frac{\sqrt{RT}}{2\pi E}
\left(1-\cos(kL) -w \sin(kL) \right) \;\;.
\label{deltados}
\end{equation}
A calculation of the global DOS from (\ref{tdoss}) yields a
wrong result without oscillation terms. Such oscillations in the
DOS and the PDOS should influence the conduction properties
of sufficiently small conductors.\\ \indent
Using Eq. (\ref{deltagreens}) one obtains for the sensitivity
\begin{equation}
\eta = \frac{w T^{2}}{h v}\:( \cos(2kx)+w\sin(2k|x|)\:) 
\label{deltasensitivity}
\end{equation}
The sensitivity for this example is a strongly oscillating
function, i. e. it contains only Friedel-like terms. Note that the
corresponding global quantity, the spatially integrated sensitivity,
\begin{equation}
\int _{-L/2}^{+L/2} dx \: \eta (x)=
\frac{w T^{2}}{4\pi E }\:( w+\sin(kL)-w\cos(kL)\:) 
\label{intesensitivity}
\end{equation}
differs strongly from the result obtained from an energy derivative of
$T$ which yields only the average value but not the oscillation
terms. Since the sensitivity is not the density of an extensive
quantity, it must be calculated by functional derivatives with
respect to the potential even in a large system.
\subsection{The local Larmor clock}
\label{larmorclock}
The Larmor clock is a system where spin-polarized electrons are
scattered by a rectangular potential barrier and an additional
perpendicular magnetic field \cite{BU2}. Outside the barrier the
magnetic field vanishes but inside it is assumed to be constant
and to point in $z$-direction. We denote the Larmor frequency by
$\omega _{L}$, and the space coordinate of the particle by $y$ rather
than by $x$. Consider electrons coming from the left side and with spin
being initially polarized in $x$-direction. Due to a Larmor precession
of the spin in the barrier, the expectation values of the spin components
depend on the time the particle spends
in the barrier. This motivates the introduction of characteristic times.
In fact, one can formally define quantities
$\tau _{x}$, $\tau _{y}$, and $\tau _{z}$ having the dimension
of a time and being associated with
the precession of each spin component $s_{x}$,
$s_{y}$, and $s_{z}$, respectively. For small $\omega_{L}$
the quantum mechanical expectation values of the spin
components of the transmitted particles are given by \cite{BU2}
\begin{eqnarray}
\langle s_{x} \rangle _{T}
& = & \frac{\hbar}{2}(1-\frac{1}{2}\omega _{L}^{2}
(\tau _{x,T})^{2})\;\;, 
\label{sx} \\
\langle s_{y} \rangle _{T}& = & -\frac{\hbar}{2}\omega _{L}\tau _{y,T}
\;\;,
\label{sy} \\
\langle s_{z} \rangle _{T}& = & \frac{\hbar}{2}\omega _{L}\tau _{z,T}
\;\;.
\label{sz}
\end{eqnarray}   
The conservation of the spin length implies $\tau _{x,T}=(\tau
_{y,T}^{2}+\tau_{z,T}^{2})^{1/2}$, i.e. only two of the
times are independent. Similarly, one can introduce times
$\tau _{x,R}$, $\tau _{y,R}$, and $\tau _{z,R}$ 
for the reflected particles. Expressions for these times have been
derived \cite{BU2} in terms of derivatives with
respect of the height of the potential barrier rather than with
respect of the particle energy. We emphasize the quantities $\tau $
defined here can be negative and, therefore, do not correspond
in general to a physical time, although they are called `Larmor
times'. Note that
the only times which are positive per definition are
$\tau _{x,T}$ and $\tau _{x,R}$. However, all of these quantities
have a clear physical meaning independent of their sign.\\ \indent
Leavens and Aers \cite{LA} discussed a local version of the Larmor
clock with an arbitrary barrier potential (as described in Sect.
\ref{section2}) and a {\em localized} magnetic field
inside the barrier. This means that $B$ is finite only in the
small interval $[x, x+dx]$  with $a<x<b$. The Larmor times are now
infinitesimal quantities proportional to the size $dx$ of the interval.
Consider particles incident from the left. It is then convenient to
introduce the following complex quantities \cite{LA}:
\begin{eqnarray}
d \tau_{t} & = & i\hbar \: \frac{\delta \ln(t)}{\delta U (x)} \: dx\;\;,
\label{tautud}\\
d\tau_{r}  & = & i\hbar \: \frac{\delta \ln(r)}{\delta U (x)} \:
dx\;\;,
\label{taurud}
\end{eqnarray}   
where $t$ and $r$ are the transmission and the reflection amplitudes,
respectively, introduced in Eq. (\ref{scatmat}). The Larmor times
are related to $d \tau _{r,t}$ by
$d \tau _{z,T} = - {\rm Im} \{ d\tau _{t} \}$,  
$d \tau _{z,R} = - {\rm Im} \{ d\tau _{r} \}$,  
$d \tau _{y,T} =   {\rm Re} \{ d\tau _{t} \}$, and  
$d \tau _{y,R} =   {\rm Re} \{ d\tau _{r} \}$. A short calculation yields
\begin{eqnarray}
d \tau_{t} & = &
\frac{h}{T}\: \left( \frac{dn_{21}}{dE}(x)
-i\eta _{21}\right) \: dx \;\;,
\label{tautu}\\
d \tau_{r} & = &
\frac{h}{R}\: \left( \frac{dn_{11}}{dE}(x)
-i\eta _{11}\right) \: dx \;\;.
\label{tauru}
\end{eqnarray}   
As it must be \cite{LA} the dwell time (\ref{dwelltime}) satisfies
\begin{equation}
d\tau _{1}= Td\tau _{t}+Rd\tau _{r} \;\;.
\label{exprob} 
\end{equation} 
Furthermore, we obtain for the Larmor times
\begin{eqnarray}
T \:d \tau_{z,T}  & = & h \: \eta _{21}(x)\: dx  \;\;,
\label{e21}\\
R \:d \tau_{z,R}  & = & h \: \eta _{11}(x)\: dx \;\;,
\label{e11}\\
T \:d \tau_{z,T}  & = & h \: \frac{dn_{21}}{dE}(x)\: dx \;\;,
\label{t21}\\
R \:d \tau_{z,R}  & = & h \: \frac{dn_{11}}{dE}(x) \: dx \;\;.
\label{t11}
\end{eqnarray}   
Similar relations hold for particles coming from the right.
The results (\ref{e21})-(\ref{t11}) connect the local PDOS with
physically well-defined quantities, which indicates the relevance of the
PDOS.\\ \indent
We re-emphasize again that it is tempting to associate the basic
PDOS $dn_{\alpha \beta}/dE$ with physically meaningful times
characterizing the tunneling process. In fact this is done in a number
of works (which do not explicitly use the terms DOS 
or even the notation used here). However, as we mentioned already
some of the PDOS can be negative which would lead to negative times.
Physical times are positive and an
{\em interpretation} of the PDOS in this direction is misleading.     
\section{Summary}
\label{section10}
In this work we discussed the decomposition of the local density
of states into partial density of states which carry the information
about the past and the future of the scattered particles. We defined
the sensitivities which describe the response of the transmission
probability to a variation of the potential.
All these quantities were defined in terms of functional 
derivatives of the scattering matrix with respect to the effective
single-particle potential which appears in the Schr\"odinger equation.
We have discussed their formal relation to the Green's function.
While the PDOS turn out to be connected to the imaginary
part of the Green's function, the sensitivity is related to its real part.
Also, their connection to characteristic times of the scattering
process was investigated and, consequently, to the absolute square of the
scattering states. Finally, we considered as simple
illustrative examples a delta-barrier in a ballistic channel and the
local Larmor clock.\\ \indent
It should be clear that the concepts introduced in this article 
apply not only to the two-channel situation
but can be generalized to many-channel scattering problems.
Furthermore, a similar point of view can very likely be developed 
even for problems with interaction for which the notion 
of an effective single-particle scattering matrix is not appropriate.
\\ \indent

{\em Acknowledgement}
This work has been supported by the Swiss National Science
Foundation. 
\section*{appendix}
We derive first the Eq. (\ref{lpdos12}). The relation
\begin{equation}
G(x,\tilde x)= \sqrt{G(x,x)G(\tilde x,\tilde x )}\:
\exp \left (\int_{\min(x,\tilde x)}^{\max(x,\tilde x)}\frac{dz}
{2\:G(z,z)}\right )
\label{app1}
\end{equation}
provided by Ref. \cite{AGG} implies for $x_{\alpha}<x<x_{\beta}$
immediately $G(x_{\alpha},x)G(x,x_{\beta})=G(x_{\alpha},x_{\beta})
G(x,x)$. It follows
\begin{equation}
\frac{dn_{12}}{dE}(x)  \; = \; i \: \frac{\hbar ^{2}v_{1}v_{2} }{4\pi}
|G(x_{1},x_{2})|^{2} \:(\:G(x,x)-G^{\ast}(x,x)\:)
\label{app2}
\end{equation}   
which is equivalent to Eq. (\ref{lpdos12}).
Next, we prove the validity of Eqs. (\ref{lpdos11}) and (\ref{lpdos22})
for $x\in \Omega _{1}$. To do this we mention first
a relation between the scattering matrix elements $s_{12}$ and
$s_{x2}$, where $s_{x2}$ connects current amplitudes
at $x_{2}$ and at $x$. Using the technique of transfer matrices and their
relation to scattering matrices, one obtains $s_{12}=s_{x2}\exp
ik(x-x_{1})$. Similarly, one shows that $s_{11}=s_{xx}\exp
2ik(x-x_{1})$ where $s_{xx}$ denotes the reflection amplitude at $x$.
With the help of the Fisher-Lee relation
(\ref{fisherlee}) one shows     
\begin{eqnarray}
\frac{dn_{22}}{dE}(x) & = & -\frac{\hbar v_{2}}{4\pi}\left( 
s_{22}^{\ast}G^{2}(x_{2},x)\:+ \: h.c \right)
\nonumber \\
& = & \frac{1}{4\pi\hbar v_{1}}\left(
i\sqrt{R}\exp (i\phi +i\phi _{a})\: T\: \exp (-2ik(x-x_{1})) \; + \;
{\rm h.c.} \right)
\nonumber \\ 
& = & -\frac{T}{2h v_{1}} \left( s_{xx}+s_{xx}^{\ast} \right)
\; = \; \frac{T}{hv_{1}}+\frac{T}{4\pi i}\left( G(x,x) -G^{\ast}(x,x)\right)
\label{app3}
\end{eqnarray}
which yields Eq. (\ref{lpdos22}). In similar way, one has
\begin{eqnarray}
\frac{dn_{11}}{dE}(x) & =& -\frac{\hbar v_{1}}{4\pi}\left( 
s_{11}^{\ast}G^{2}(x_{1},x)\:+ \: h.c \right)
\nonumber \\
& = & -\frac{\hbar v_{1}}{4\pi}\left( 
s_{11}^{\ast}\frac{G^{2}(x_{1},x)G^{2}(x_{2},x)}{G^{2}(x_{2},x)}
\:+ \: h.c \right)
\nonumber \\
& =&  -\frac{\hbar v_{1}}{4\pi}\left( 
s_{11}^{\ast}\frac{G^{2}(x_{1},x_{2})G^{2}(x,x}{G^{2}(x_{2},x)}
\:+ \: h.c \right)
\nonumber \\
& = & -\frac{\hbar v_{1}}{4\pi}\left( 
s_{xx}^{\ast}G^{2}(x,x) \:+ \: h.c \right)
\; = \; \frac{R+1}{2hv_{1}}\left( s_{xx} + s_{xx}^{\ast} \right)
\label{app4}
\end{eqnarray}
which yields Eq. (\ref{lpdos11}). Finally, the expression
(\ref{sens1}) for the sensitivity follows from
\begin{equation}
\frac{\delta |s_{12}|}{\delta U(x)} = \hbar ^{2} v_{1}v_{2}\left(
|G(x_{1},x_{2})|^{2} G^{\ast}(x,x) \; + \; {\rm h.c.} \right)\;\;.
\label{app5}
\end{equation} 

\newpage
\begin{figure}
\caption{One-dimensional scattering problem described in the text.
  The dashed and dotted curves belong to Fermi-energies
  associated with a transparent and an opaque barrier, respectively.}
\label{fig1}
\end{figure}
\begin{figure}
\caption{Classical phase-space plot of the scattering region in
  Fig. \protect\ref{fig1}. The WKB density of states is obtained
  from the phase-space area $d\Phi$ (grey region)
  between two trajectories (thin lines) of energy difference $dE$.
  The dashed arrows indicate tunneling.}
\label{fig2}
\end{figure}
\begin{figure}
\caption{Partial densities of states $dn_{11}/dE$ (dashed),
  $dn_{12}/dE$ (dotted), $dn_{22}/dE$ (dashed-dotted), and
  $dn/dE$ (solid) for the delta barrier with $T=0.8$.}
\label{fig3}
\end{figure}
\begin{figure}
\caption{Injectivity (or absolute square of the scattering
  wave-function) $d \overline{n}_{1}/dE$ for the delta barrier
  with $T=0.8$.}
\label{fig4}
\end{figure}
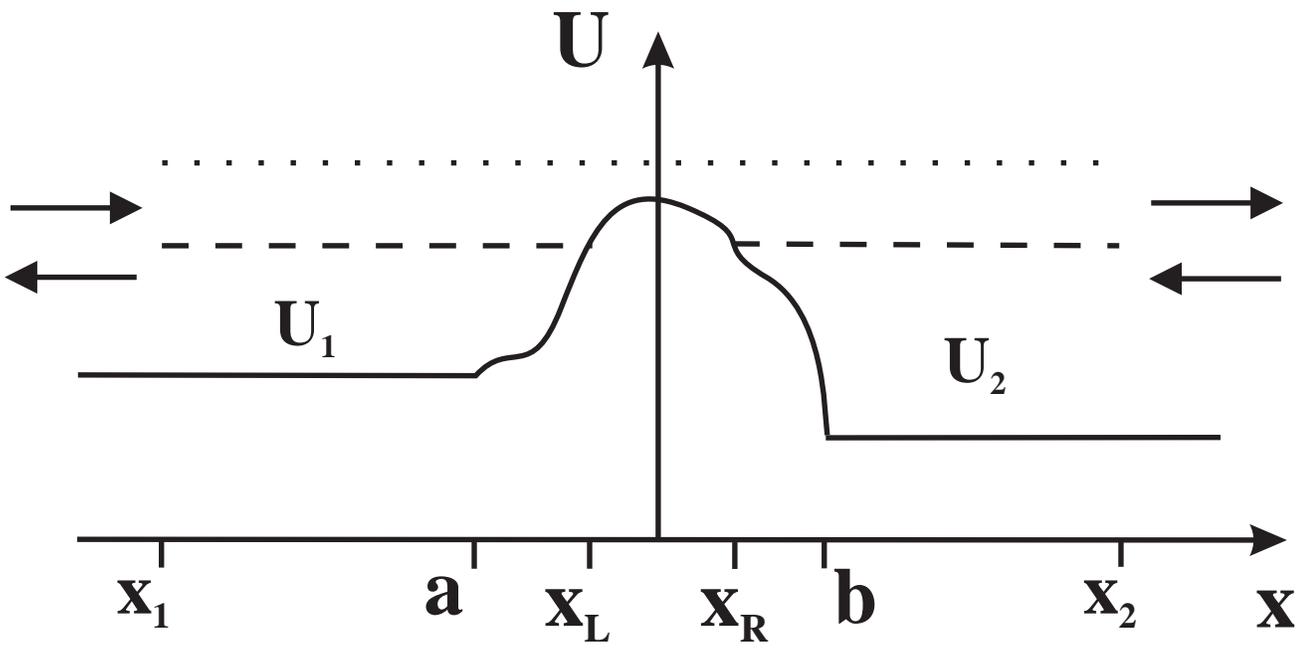
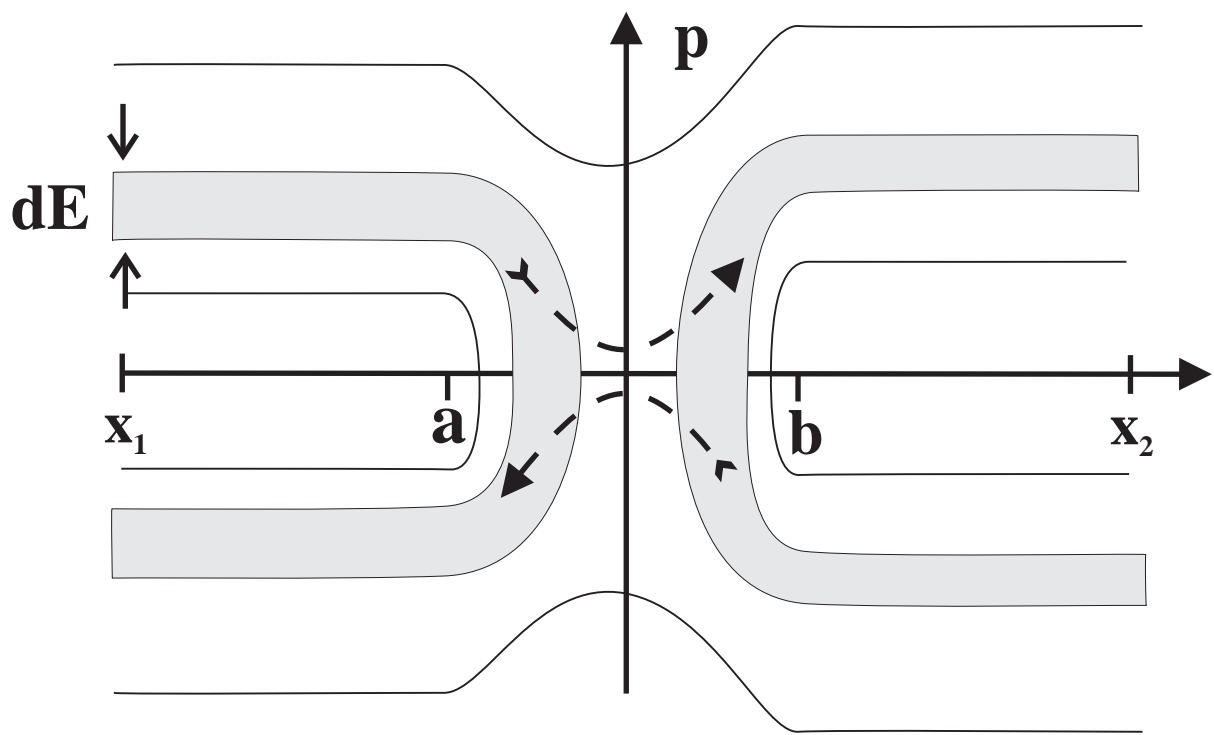
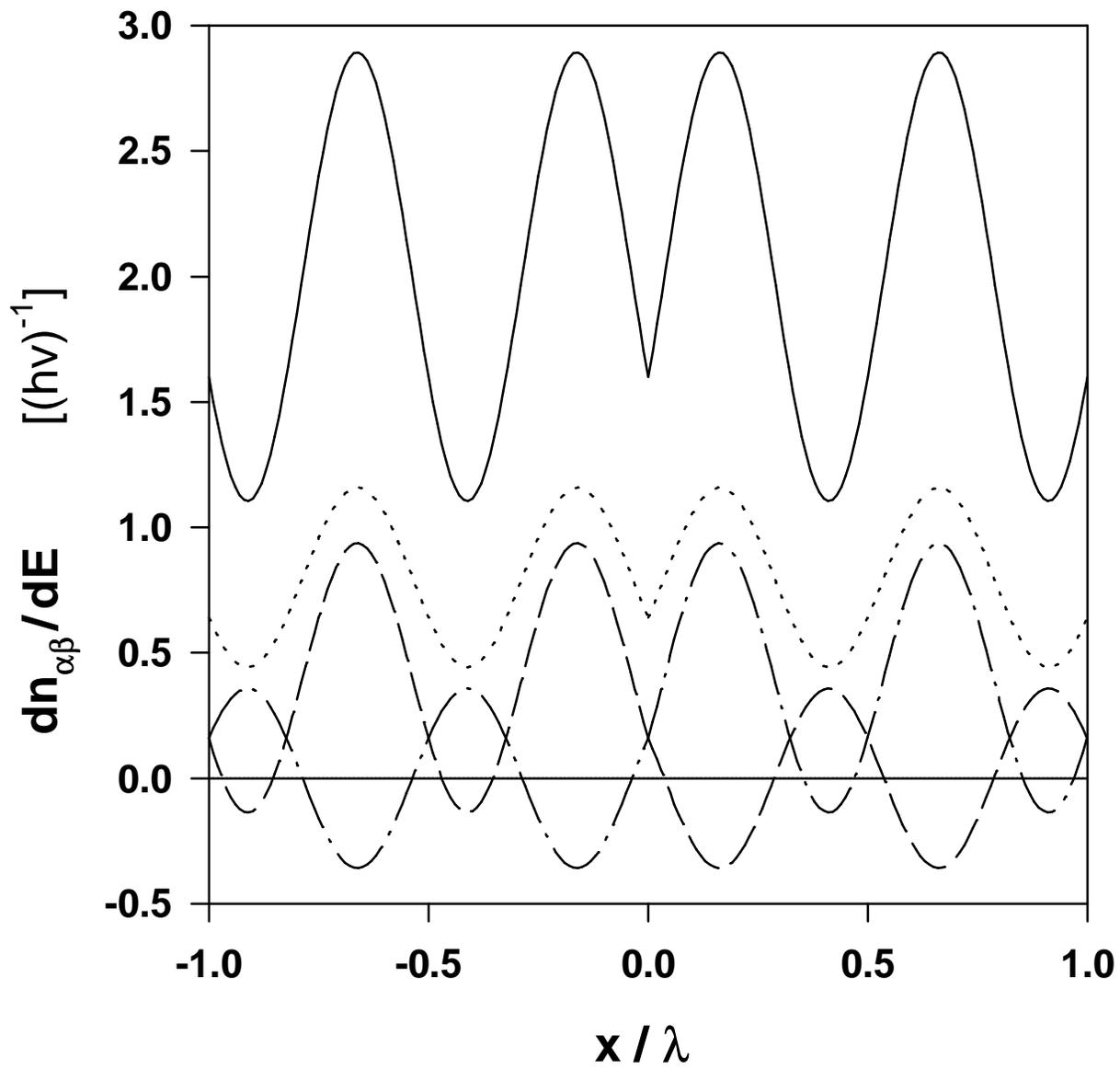
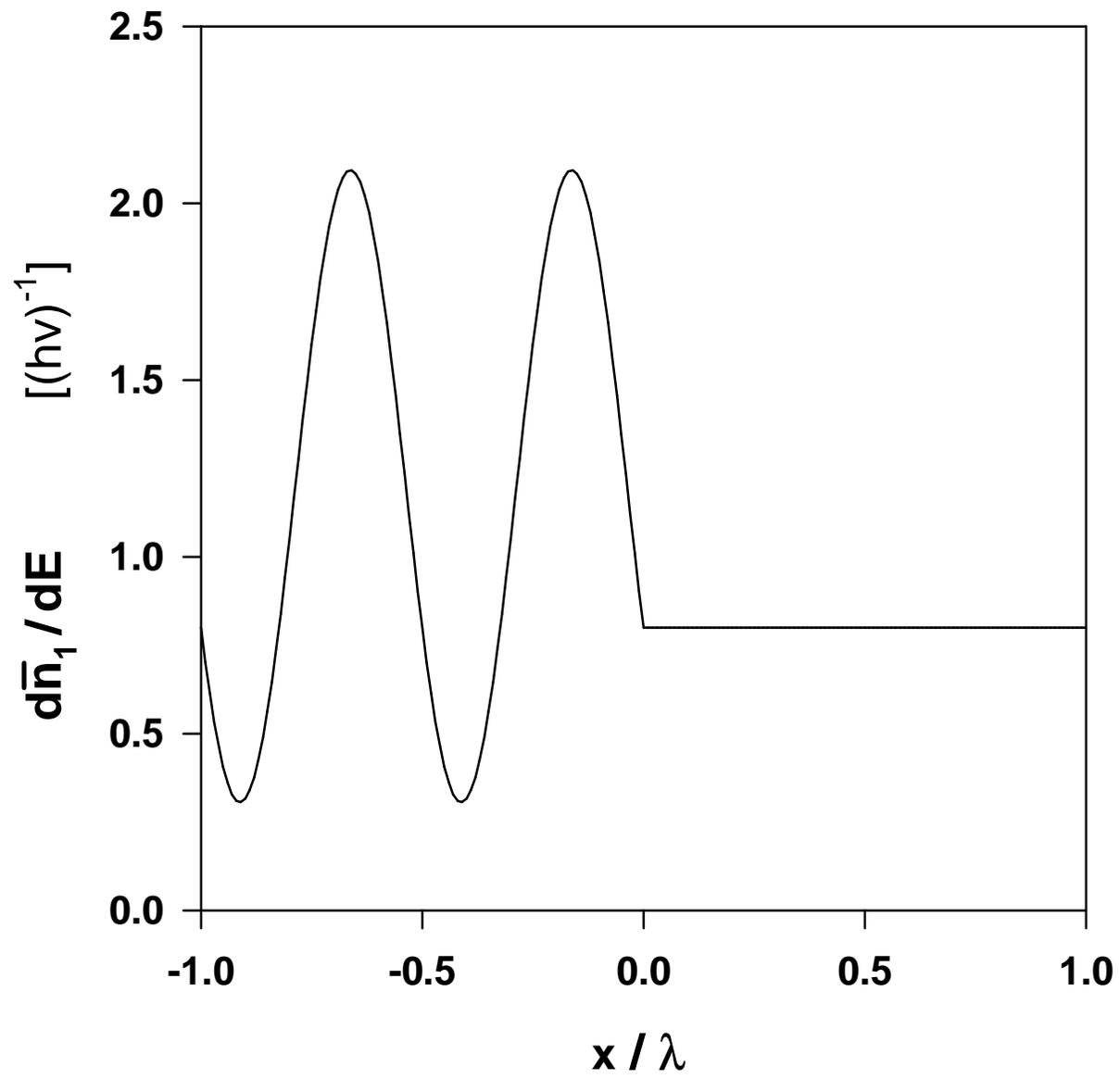
\end{document}